\documentclass[10pt,twocolumn,letterpaper]{article}

\usepackage{cvpr}
\usepackage{times}
\usepackage{epsfig}
\usepackage{graphicx}
\usepackage{amsmath}
\usepackage{amssymb}
\usepackage{mathrsfs}
\usepackage[table,xcdraw]{xcolor}

\usepackage[pagebackref=true,breaklinks=true,letterpaper=true,colorlinks,bookmarks=false]{hyperref}

\cvprfinalcopy 

\def\cvprPaperID{9} 

\ifcvprfinal\fi
\begin{document}

\title{S2A: Wasserstein GAN with Spatio-Spectral Laplacian Attention for Multi-Spectral Band Synthesis}

\author{Litu Rout \\
\and Indranil Misra \and S Manthira Moorthi \and Debajyoti Dhar \and Signal and Image Processing Group\\
Space Applications Centre\\
Indian Space Research Organisation\\
{\tt\small (lr, indranil, smmoorthi, deb)@sac.isro.gov.in}
}
\maketitle

\begin{abstract}
   Intersection of adversarial learning and satellite image processing is an emerging field in remote sensing. In this study, we intend to address synthesis of high resolution multi-spectral satellite imagery using adversarial learning. Guided by the discovery of attention mechanism, we regulate the process of band synthesis through spatio-spectral Laplacian attention. Further, we use Wasserstein GAN with gradient penalty norm to improve training and stability of adversarial learning. In this regard, we introduce a new cost function for the discriminator based on spatial attention and domain adaptation loss. We critically analyze the qualitative and quantitative results compared with state-of-the-art methods using widely adopted evaluation metrics. Our experiments on datasets of three different sensors, namely LISS-3, LISS-4, and WorldView-2 show that attention learning performs favorably against state-of-the-art methods. Using the proposed method we provide an additional data product in consistent with existing high resolution bands. Furthermore,  we synthesize over 4000 high resolution scenes covering various terrains to analyze scientific fidelity. At the end, we demonstrate plausible large scale real world applications of the synthesized band\footnote{Accepted for publication at Computer Vision and Pattern Recognition (CVPR) Workshop on Large Scale Computer Vision for Remote Sensing Imagery.}.
\end{abstract}

\section{Introduction}
\begin{figure}[t]
	\centering
	\includegraphics[width=0.9\columnwidth]{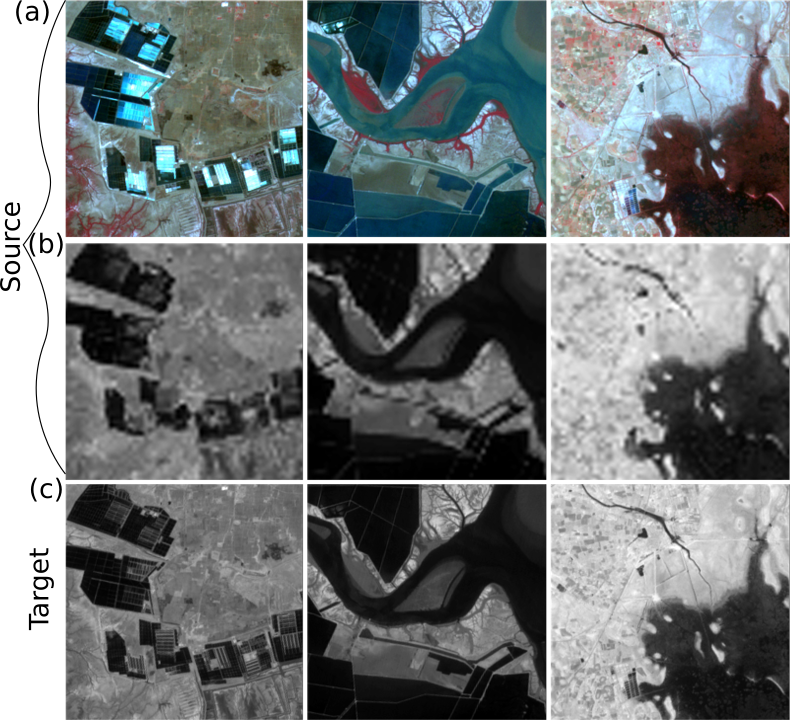}
	\caption{Paired training data. Source domain consists of (a) high resolution NIR (R), R (G) and G (B), and (b) a coarser resolution SWIR band. Target domain contains (c) corresponding high resolution SWIR band.}
	\label{pair}
\end{figure}

Attention learning is a human vision inspired algorithm that automatically attends to relevant attributes of an object. Despite its remarkable progress~\cite{vaswani2017attention,bahdanau2014neural,emami2019spa,chen2016attention,xu2015show}, necessary attention from remote sensing community has not been paid towards this particular line of research. In this study, we intend to take a step in this direction and explore attention in multi-spectral super-resolution. To make the task relatively more tractable, we formulate the ill-posed super-resolution problem as multi-spectral band synthesis. As shown in Figure~\ref{pair}, we aim to synthesize a high resolution band provided its coarser resolution band and existing high resolution multi-spectral bands.

\begin{figure*}[t]
	\centering
	\includegraphics[width=0.85\textwidth]{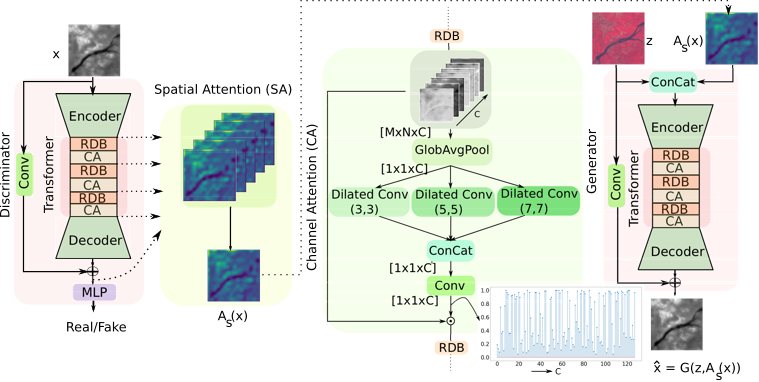}
	\caption{Overall pipeline of the proposed methodology. Both generator and discriminator share similar architecture including number of RDBs~\cite{rout2020alert}. The Multi-Layer Perceptron (MLP) (here, 3 layers) maps disentangled features to real/fake class.}
	\label{pipe}
\end{figure*}

Single image super-resolution is a widely explored field in computer vision. Recent advances in deep neural networks show compelling improvement over conventionally driven approaches on diverse datasets~\cite{yang2010image,arbelaez2010contour,bevilacqua2012low,huang2015single}. Among the early works, Dong \etal~\cite{dong2015image} proposed SRCNN, which is the first successful attempt towards employing Convolutional Neural Networks (CNN) in super-resolution. Thereafter several innovative methods~\cite{Kim_2016_CVPR,Shi_2016_CVPR,Lai_2017_CVPR,Zhang_2018_CVPR} have emerged over the years that provided better context and generalizable representation. Most recently, Anwar \etal~\cite{anwar2019densely} introduced Densely Residual Laplacian Network (DRLN) that achieved state-of-the-art results on benchmark datasets. 

Similar to supervised learning, adversarial learning based super-resolution has obtained impressive results on multitude of tasks. After the first pioneering work on single image Super-Resolution using Generative Adversarial Networks (SRGAN)~\cite{Ledig_2017_CVPR}, many derivatives have emerged~\cite{Ledig_2017_CVPR,sajjadi2017enhancenet,wang2018esrgan}. Enhanced Super-Resolution using Generative Adversarial Networks (ESRGAN)~\cite{wang2018esrgan} employed a relativistic discriminator~\cite{jolicoeur2018relativistic} along with global residual connections to allow better gradient flow. Despite inferior Peak Signal to Noise Ratio (PSNR), the adversarial learning based methods are shown to achieve higher visual perceptual quality as reported in copious literature~\cite{Ledig_2017_CVPR,sajjadi2017enhancenet,wang2018esrgan,park2018srfeat}.

In remote sensing community, deep learning based single image super-resolution is a rapidly evolving field. Huang \etal~\cite{huang2017single} used residual learning modules in remote sensing image super-resolution. Luo \etal~\cite{luo2017video} designed CNN based video satellite image super-resolution. Lei \etal~\cite{lei2017super} combined global and local features to introduce high resolution features in a coarser resolution remote sensing image. Beaulieu \etal~\cite{beaulieu2018deep} applied CNN to super-resolve Sentinel-2 imagery. Recently, Mario \etal~\cite{haut2018new} proposed an unsupervised approach to super-resolve remote sensing images.

Among high resolution band synthesis approaches, Lanaras \etal~\cite{lanaras2018super} developed a super-resolution framework that exploits resolution invariant nature of deep neural networks. Further, Rout \etal~\cite{rout2019deepswir} used global and local residual learning to fuse spectral and spatial characteristics of concurrent multi-resolution bands in band synthesis. Rangnekar \etal~\cite{rangnekar2017aerial} proposed adversarial learning with $L_1$-penalty to improve the spectral resolution of aerial imagery. More recently, L. Rout~\cite{rout2020alert} designed an adversarial learning mechanism with expert regularization using Wasserstein GAN~\cite{gulrajani2017improved} to synthesize a missing band in multi-spectral images. The authors of~\cite{rout2020alert} demonstrated efficient gradient flow due to Residual Dense Blocks (RDBs). While resolution invariant property~\cite{lanaras2018super} works well in certain situations, we argue that it fails to \textit{extrapolate} the spatial resolution with similar accuracy. In addition, due to significant dependency on coarser band,  it attributes towards computational bottleneck as it requires precise cross-sensor registration during operation. For this reason, we build our framework on synthesizing SWIR band using existing concurrent high resolution bands. The proposed method does not directly take coarser resolution band as input. It only requires a glimpse of the coarse band just to identify relevant parts of the image, which will be discussed in detail later. Different from~\cite{rout2020alert}, we use spatial and Laplacian spectral channel attention along with two newly introduced cost functions in adversarial learning. Figure~\ref{pipe} pictorially depicts our overall framework.

The rest of the paper is organized as follows. In Section~\ref{relwork}, we briefly discuss about the prior methods relevant to this study. We describe the detailed methodology in Section~\ref{method}, and experimental details with analysis in Section~\ref{exp}. Finally, we draw concluding remarks in Section~\ref{conc}.

\section{Related Work}
\label{relwork}
\subsection{Image-to-Image Translation}
\label{transln}
GANs have been rigorously studied both from theoretical and application perspective in the last few years. Adversarial learning introduced a concept of continuously evolving objective function that provides an edge over fixed objectives. In certain applications~\cite{zhu2017unpaired}, where collection of paired data is very costly, GANs learn a loss that adapts to the data. In this line of work, Isola \etal~\cite{isola2017image} used Conditional GANs, known as (cGAN) in image-to-image translation. While cGAN learns from paired images, Zhu \etal~\cite{zhu2017unpaired} designed CycleGAN that learns from unpaired images. Sangwoo \etal~\cite{mo2018instagan} introduced context preserving loss by using semantic segmentation labels. Though it preserves background during translation, it has a major limitation for new applications where semantic labels are not available. In a loose sense, our work can be characterized as image-to-image translation where source domain contains composites of existing low resolution bands and target domain, the \textit{paired} target band. Our work is more similar to~\cite{isola2017image} as we condition the synthesis process on existing concurrent bands. Since generation of any band is obviously not acceptable, we follow the common practice of choosing cGAN. Different from these prior works, we use global and local residual learning with spatio-spectral attention in a WGAN framework. In particular, we explore the efficacy of these methods in multi-spectral satellite images, which is the primary focus of this study.

\subsection{Attention Learning}
\label{attn}
Attention learning is a mechanism to automatically attend to relevant parts of an image while performing certain tasks. Inspired by human visual system, attention mechanism focuses on salient attributes of an object. Attention adds an extra layer to the interpretability of deep neural networks. It improves the performance in various tasks~\cite{bahdanau2014neural,emami2019spa,chen2016attention,xu2015show} encouraging further investigation in remote sensing. Zagoruyko \etal~\cite{zagoruyko2016paying} demonstrated that the absolute value of activation of a hidden neuron is proportional to the importance of that neuron in performing a desired task. Mnih \etal~\cite{mnih2014recurrent} proposed a recursive visual attention model that processed sequence of regions at a high resolution. Wang \etal~\cite{wang2017residual} used residual attention for image classification.

Among generative models, Zhang \etal~\cite{zhang2018self} designed Self-Attention GAN (SAGAN) that incorporated attention in the process of image generation. Emami \etal~\cite{emami2019spa} showed the benefits of using spatial attention from the discriminator. In DRLN~\cite{anwar2019densely}, the authors added an extra layer in channel attention by introducing Laplacian pyramid. Despite increasing popularity of attention learning, necessary attention has not been paid explicitly towards this line of research in the remote sensing community. As a step towards achieving this goal, Bastidas \etal~\cite{bastidas2019channel} proposed Channel Attention Network (CAN) that used soft attention in multi-spectral semantic segmentation. In this study, we take a step further to explore the plausible usage of spatial and Laplacian spectral (channel) attention in band synthesis. We incorporate these attention modules in WGAN with gradient penalty. In addition, we introduce spatial attention and domain adaptation loss for efficient learning.

\section{Methodology}
\label{method}
Following the notations from~\cite{rout2020alert}, let $z_1 \sim \mathbb{P}_G$, $z_2 \sim \mathbb{P}_R$, $z_3 \sim \mathbb{P}_{NIR}$, and $x \sim \mathbb{P}_{SWIR}$, where $\mathbb{P}_G$, $\mathbb{P}_R$, $\mathbb{P}_{NIR}$, and $\mathbb{P}_{SWIR}$ represent the distribution of $G, R, NIR$, and $SWIR$, respectively. The source domain consists of samples from the joint distribution $\mathbb{P}_{S}(z_1,z_2,z_3)$, where $z_i \in \mathbb{R}^{M\times N}$, $i=1,2,3$. The generator, $G$ operates on conditional input, $z \sim \mathbb{P}_S$ and spatial attention map, $A_s \in \mathbb{R}^{M\times N}$ from the discriminator, $D$. Let $\hat{x} \sim \mathbb{P}_{\hat{x}}$, where $\hat{x}=G(z,A_s)$ and $\mathbb{P}_{\hat{x}}$ denotes the generator distribution. The discriminator classifies $x \sim \mathbb{P}_x$ and $\hat{x}$ to real and fake categories, respectively. Here, $\mathbb{P}_x$ denotes the target distribution, $\mathbb{P}_{SWIR}$. The expert system has access to samples from target domain that correspond to the physical landscape of identical samples in source domain. Let $y \sim \mathbb{P}_{y}$ represents the corresponding target sample in SWIR band. To study the impact of attention mechanisms, we build on top of the baseline architecture as developed in~\cite{rout2020alert}. The overall pipeline is shown in Figure~\ref{pipe}.

\subsection{Adversarial Loss}
After the discovery of GANs by Goodfellow \etal~\cite{goodfellow2014generative}, several variants of adversarial networks have been proposed~\cite{mirza2014conditional,radford2015unsupervised,arjovsky2017wasserstein,gulrajani2017improved}. In this study, we focus on Wasserstein GANs with gradient penalty due to its ability to capture difficult-to-learn latent patterns~\cite{gulrajani2017improved}. Thus, the min-max objective function of WGAN+GP adapted to current setting is given by
\begin{equation}
	\begin{split}
	\min_{G} \max_{D} &~\mathbb{E}_{x \sim \mathbb{P}_{x}}\left [D\left ( x \right )  \right ]-\mathbb{E}_{\hat{x} \sim \mathbb{P}_{\hat{x}}}\left [D\left ( \hat{x} \right ) \right ] \\
	& - \lambda~\mathbb{E}_{\tilde{x} \sim \mathbb{P}_{\tilde{x}}}\left [ \left ( \left \| \nabla_{\tilde{x}}D\left ( \tilde{x} \right ) \right \|_2 -1 \right )^2 \right ],
	\end{split}
\end{equation}
where $\mathbb{P}_{\tilde{x}}$ denotes the distribution of samples along the line of samples from $\mathbb{P}_x$ and $\mathbb{P}_{\hat{x}}$. In~\cite{gulrajani2017improved}, the authors argue that such gradient penalty is sufficient to maintain stability during training. Also, it broadens the hypothesis space that can be approximated by this estimator.

\subsection{Spatial Attention from Discriminator}
The spatial attention map from the discriminator ensures that the generator focuses on relevant parts of the input images during domain-to-domain translation. Since discriminator classifies images into real or fake categories, evidently it captures discriminative features in latent space. Thus, identification of these regions serves as spatial attention that can assist a generator to focus its attention.

We follow activation-based attention transfer as described by Zagoruyko \etal~\cite{zagoruyko2016paying}. Similar to transferring attention of a teacher CNN to a student CNN, we transfer knowledge of the discriminator to the generator through spatial attention maps. In~\cite{emami2019spa}, the authors use normalized spatial attention maps from the discriminator to transfer domain specific features. However, in the context of super-resolution, or band synthesis such a straight forward implementation might not be adequate. A main reason could be the absence of high-resolution bands in the target domain. For this reason, we introduce a notion of spatial attention loss and cross-resolution attention transfer during training.

Formally, $D$ consists of two branches: a functional branch to classify an image as real or fake, $D_{rf}:\mathbb{R}^{M\times N}\rightarrow \mathbb{R}$ and a computational branch to estimate its spatial attention, $D_{s}:\mathbb{R}^{M\times N}\rightarrow [0,1]^{M\times N}$. For $K$ RDBs and $C$ spectral channels in the output of each RDB, let $A_i\in\mathbb{R}^{M\times N\times C}$ denote the activation maps after $i^{th}$ RDB. Since different layers focus on different features, we extract $K$ attention maps from various layers in the latent space. Finally, we estimate the attention coefficients as:
\begin{equation}
\begin{split}
    A_s(x) &= \mathcal{N}\left ( D_s(x) \right ),\\
    D_s(x) &= \sum_{i=1}^{K}\mathcal{N}\left ( \sum_{j=1}^{C} \left | A_{ij} (x) \right | \right ),
\end{split}
\label{sp_attn_eq}
\end{equation}
where $\mathcal{N}(.)$ normalizes inputs to [0,1] range. In super-resolution, we usually do not have high resolution samples in the target domain unlike image-to-image translation. For this reason, we use upsampled coarse resolution image to compute the attention maps. Also, the geometric fidelity and band-to-band registration are ensured by taking into account the importance of each neuron at every pixel location. This is asserted by global skip connection and normalization of activation maps in the latent space. The attention map and conditional input from source domain are then passed to the generator.

\subsubsection{Spatial Attention Loss}
In addition, we introduce an additional loss to ensure that discriminator attends to similar regions in both real and fake images while classifying them into real and fake class, respectively. This penalizes the discriminator for attending to different locations in real and fake images. Thereby, the discriminator learns to transfer both low and high level semantics of target band. We define the spatial attention loss as
\begin{equation}
    \begin{split}
        \mathscr{L}_{sa} = \mathbb{E}_{\hat{x}\sim \mathbb{P}_{\hat{x}},y\sim\mathbb{P}_y}\left [ \left \| A_s(\hat{x}) - A_s(y) \right \|^2_2 \right ].
    \end{split}
\end{equation}

\subsubsection{Domain Adaptation Loss}
 Due to the absence of high resolution band in testing phase, the discriminator is expected to predict relevant parts based on existing coarse resolution band. By domain adaptation loss, we enforce the discriminator to mimic the spatial attention map of actual high resolution band provided an upsampled low resolution band. Thus, the discriminator captures domain specific features so as to sharpen the spatial attention map which was obtained using blurry target band. The domain adaptation loss is defined by
 \begin{equation}
    \begin{split}
        \mathscr{L}_{da} = \mathbb{E}_{\tilde{y}\sim \mathbb{P}_{\tilde{y}},y\sim\mathbb{P}_y}\left [ \left \| A_s(\tilde{y}) - A_s(y) \right \|^2_2 \right ],
    \end{split}
\end{equation}
where $\mathbb{P}_{\tilde{y}}$ denotes the distribution of upsampled coarse resolution band.

\subsection{Spectral Attention from Generator}
To equip the generator with tools so that it can attend to relevant spectral channels, we employ Laplacian channel attention after every RDB. The sparse channel coefficients learned by generator suggest that spectral attention unit automatically eliminates the noisy spectral channels in latent space. For $K$ RDBs and $C$ spectral channels in each RDB, let $F_i\in\mathbb{R}^{M\times N\times C}$ denote the feature maps after $i^{th}$ RDB. First, we apply global average pooling to reduce spatial dimension while preserving spectral dimension, i.e., $M=1$ and $N=1$. By convolution with kernel size (3,3), padding (1,1), and dilation rate of 3, 5 and 7, we construct Laplacian pyramid to reduce channel dimension by a factor of 16.   We then concatenate the pyramidal features in channel dimension, and apply convolution with kernel size (3,3) and padding (1,1) to generate spectral channel attention coefficients, $A_c (F_i) \in \mathbb{R}^{1\times 1\times C}$. Thus, the importance of each channel in latent space is estimated automatically for efficient band-to-band synthesis. Finally, each feature map is modulated by $F_i \odot A_c(F_i)$ before being passed as an input to the next RDB. Figure~\ref{pipe} shows spectral attention as a part of the overall framework.

\subsubsection{Pixel Loss}
To make images more realistic in the target domain, a generator can synthesize many images satisfying this criteria. However, these images may not represent the physical landscape pertaining to the input image in source domain. Therefore, to ensure faithful generation, we regularize the objective function of generator with a pixel loss. Further, we pretrain the generator for few epochs (here, 2) using pixel loss before adversarial training. This offers faster convergence and minimal empirical risk as argued in~\cite{rout2020alert}. The pixel loss used in this framework is given by
\begin{equation}
\begin{split}
	\mathscr{L}_p = \mathbb{E}_{z\sim \mathbb{P}_s, \tilde{y} \sim \mathbb{P}_{\tilde{y}},y\sim \mathbb{P}_y}\left [ \left \| G\left ( z,A_s(\tilde{y}) \right ) -y \right \|_2^2 \right ].
\end{split}
\end{equation}
It is worth mentioning that the generator learns to focus on relevant parts of the source image, $z$ by attending to upsampled coarse resolution band, $A_s(\tilde{y})$. By domain adaptation loss, the discriminator makes $A_s(\tilde{y})$ close to $A_s(y)$, which consequently improves the performance of the generator.

\subsection{Total Loss}
Here, we consolidate the min-max objectives of generator and discriminator in adversarial setting. After incorporating the aforementioned losses, the final objective fuction of the discriminator becomes
\begin{equation}
\begin{split}
	\min_D \mathbb{E}_{\hat{x}\sim\mathbb{P}_{\hat{x}}}\left [ D\left ( \hat{x} \right ) \right ] & - \mathbb{E}_{x\sim\mathbb{P}_{x}}\left [ D\left ( x \right ) \right ] \\ &+ \lambda_{gp}\mathbb{E}_{\tilde{x} \sim \mathbb{P}_{\tilde{x}}}\left [ \left ( \left \| \nabla_{\tilde{x}}D\left ( \tilde{x} \right ) \right \|_2 -1 \right )^2 \right ] \\
& + \lambda_{sa}\mathscr{L}_{sa}+ \lambda_{da}\mathscr{L}_{da},
\end{split}
\end{equation}
where $\lambda_{gp}, \lambda_{sa}$, and $\lambda_{da}$ represent the weights assigned to gradient penalty, spatial attention, and domain adaptation loss, respectively. Similarly, the objective function of the generator is given by
\begin{equation}
\begin{split}
	\min_G & - \mathbb{E}_{z\sim \mathbb{P}_s, \tilde{y} \sim \mathbb{P}_{\tilde{y}}}\left [ D\left ( G\left ( z,A_s(\tilde{y}) \right ) \right ) \right ] \\ &+ \lambda_p \mathbb{E}_{z\sim \mathbb{P}_s, \tilde{y} \sim \mathbb{P}_{\tilde{y}},y\sim \mathbb{P}_y}\left [ \left \| G\left ( z,A_s(\tilde{y}) \right ) -y \right \|_2^2 \right ],
\end{split}
\end{equation}
where $\lambda_p$ represents the weight assigned to pixel loss.

\section{Experiments}
\label{exp}
In this section, we provide detailed description of our experiments to critically analyze the efficacy of proposed methodology.

\begin{figure}[t]
	\centering
	\includegraphics[width=0.7\columnwidth]{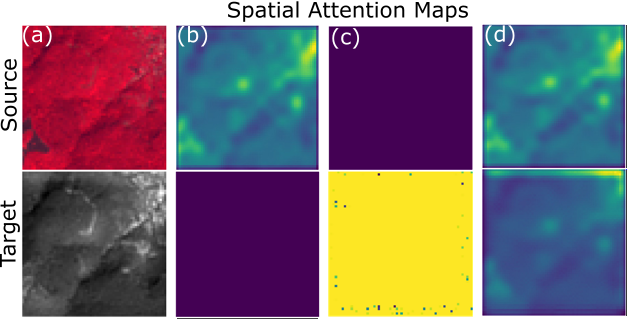}
\caption{Spatial attention maps from various discriminators. (a) Paired Images in source and target domain. Attention maps from (b) S2A-v1 (c) S2A-v2, and (d) S2A-v3. Encoder and final attention maps are shown in upper and lower row, respectively.}
	\label{sp_attn}
\end{figure}

\subsection{Datasets and Study Area}
We use Indian Remote Sensing (IRS) satellite Resourcesat-2A (R2A) and WorldView-2 (WV-2) data in the development process. The training data consists of 32193 crops of size (64,64) which include various signatures, such as vegetation, inland water, ocean, land, mountain, road, urban area, and cloud. We use 5682 crops for 1-fold cross validation during training. In the initial phase, we test on Linear Imaging and Self-Scanning Sensor-3 (LISS-3) onboard Resourcesat-2A, which has a Ground Sampling Distance (GSD) 24m. In this case, the testing data contains 19050 crops which include most of the aforementioned features. Also, the model is tested on multi-sensor and multi-resolution datasets of LISS-4 (GSD=6m) and WV-2 (GSD=1.84m) despite fully trained on LISS-3. We test on 101023  crops of LISS-4 and 16510 crops of WV-2. For seamless band synthesis, we use overlapping patches of stride (16,16) and stitch them together using Gaussian feather mosaic~\cite{rout2019deepswir}. Though the model is trained on LISS-3 over Indian territory, we test on a completely isolated physical landscape, Washington in order to study its ability to generalize. Overall, we analyze its performance on multi-temporal, multi-resolution, and multi-sensor datasets.

\begin{figure}[t]
	\centering
	\includegraphics[width=\columnwidth]{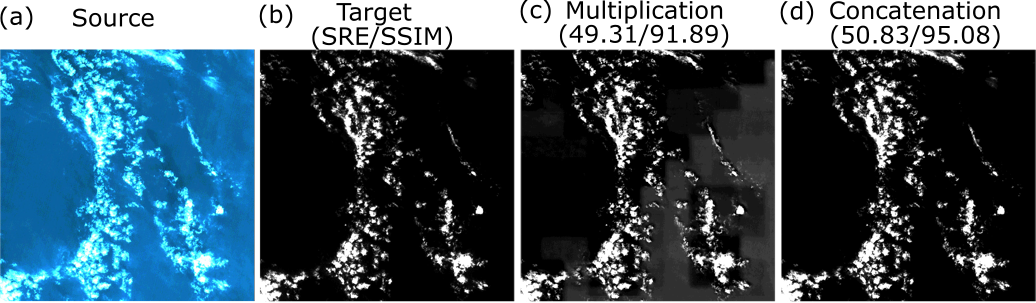}
\caption{Comparison between element-wise multiplication and concatenation of spatial attention map (contrast stretched).}
	\label{mul_con}
\end{figure}

\begin{figure*}[t]
	\centering
	\includegraphics[width=0.75\textwidth]{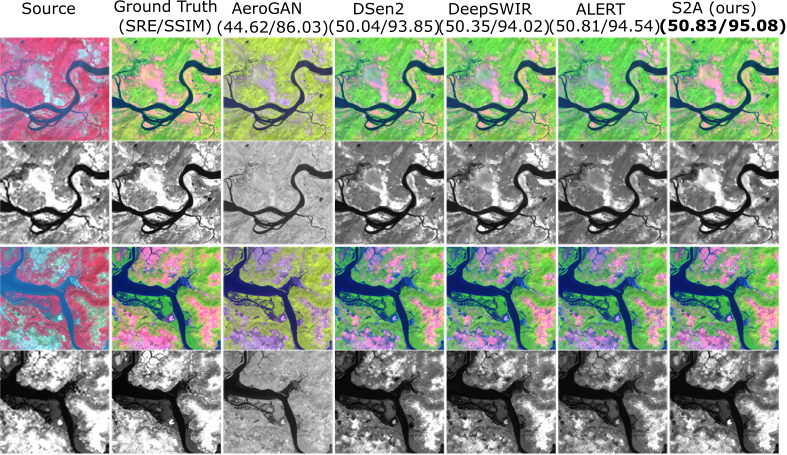}	
	\caption{Analysis of multi-temporal LISS-3 SWIR band (better viewed at 200\%). Band combination used in visualization: SWIR (R), NIR (G), R (B), and corresponding SWIR (gray). Best approach is highlighted in bold font.}
	\label{mul_temp}
\end{figure*}

\begin{figure}[t]
	\centering
	\includegraphics[width=0.7\columnwidth]{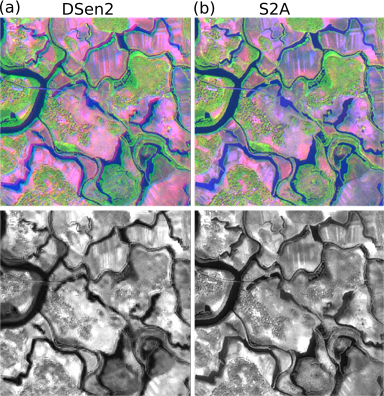}
	\caption{Analysis of multi-resolution (4x) LISS-4 SWIR band synthesis (better viewed at 300\%). Band combination: SWIR (R), NIR (G), R (B), and corresponding SWIR (gray).}
	\label{mul_res}
\end{figure}

\subsection{Implementation Details}
Here, we provide necessary details to reproduce the results reported in this paper. All the experiments are conducted under identical setup. Both generator and discriminator share similar architecture. The encoder and decoder consist of two convolution layers each. We use (3,3) convolutional kernels everywhere except in global skip connection, where (1,1) kernel is used. In channel attention module, the final convolution layer uses sigmoid activation. The MLP in discriminator has 3 linear layers with leaky ReLU activation. The output linear layer in MLP does not use any activation. There are 6 RDBs~\cite{rout2020alert} in the transformer block of both generator and discriminator. Each RDB uses 128 convolutional kernels, i.e., $C = 128$ and ReLU activations. We use ADAM optimizer with fixed learning rate 0.0001. During adversarial learning, we update critic once for every single update in the generator. We set $\lambda_{gp} = 10, \lambda_{sa} = 0.1, \lambda_{da} = 0.1$, and $\lambda_{p} = 100$. The entire framework is developed using PyTorch.

\subsection{Ablation Study}
In this section, we analyze various spatial attention configurations to address the problem under study. First, we focus on the spatial attention maps from discriminator. As shown in Figure~\ref{pipe}, we extract individual attention maps after encoder, RDBs, and decoder of the discriminator. For conciseness, we focus on the final attention map and the map after encoder. In S2A-v1 and S2A-v2, the attention maps are computed by equation~(\ref{sp_attn_eq1}) and equation~(\ref{sp_attn_eq2}), respectively. Here, $\sigma(.)$ represents sigmoid activation. 
\begin{equation}
\begin{split}
    A_s(x) = \sigma\left ( \sum_{i=1}^{K}\left ( \sum_{j=1}^{C} \left | A_{ij} (x) \right | \right ) \right )
\end{split}
\label{sp_attn_eq1}
\end{equation}

\begin{equation}
\begin{split}
    A_s(x) = \sigma\left ( \sum_{i=1}^{K} \sigma \left (\sum_{j=1}^{C} \left | A_{ij} (x) \right | \right ) \right )
\end{split}
\label{sp_attn_eq2}
\end{equation}
The proposed attention maps are computed using equation~(\ref{sp_attn_eq}). As shown in Figure~\ref{sp_attn}, the final attention map of S2A-v1 is saturated even though it attends to relevant parts in the latent space. Also, we observe similar phenomena in S2A-v2 where both latent and final attention maps lack identifiable relevant parts. Contrary to that S2A-v3 captures relevant information in all the attention maps. To prevent attention maps from saturating while attending to essential attributes, we normalize individual feature maps in S2A-v3 by \textit{minimum subtraction} and \textit{maximum division}, which is denoted by operator $\mathcal{N}$ in equation~(\ref{sp_attn_eq}).

Further, we study the impact of multiplying spatial attention map with source image as opposed to concatenating them in the context of multi-spectral band synthesis. As shown in Figure~\ref{mul_con}, concatenation results in superior image quality in terms of SRE~\cite{lanaras2018super} and SSIM as compared to multiplication. The underlying hypothesis is that due to dominance of brighter targets, such as cloud, the element-wise multiplication based spatial attention latches on to these objects in a patch. For this reason, we observe blocky artifacts around high reflecting targets. On the other hand, the concatenation based spatial attention does not suffer from this issue as the original input is still preserved in one of the dimensions. This is evident from Figure~\ref{mul_con} where each neighborhood of cloudy targets incurs blocky artifacts. 

\subsection{Quantitative Analysis}
Here, we quantitatively analyze the quality of synthesized band. Following the state-of-the-art methods in multi-spectral band synthesis~\cite{rangnekar2017aerial,lanaras2018super,rout2019deepswir,rout2020alert}, we use RMSE, SSIM, SRE~\cite{lanaras2018super}, PSNR, and SAM~\cite{yuhas1992discrimination} as our evaluation metrics. As reported by Lanaras \etal~\cite{lanaras2018super}, SRE is preferred over PSNR to analyse the quality of a satellite image due to its high dynamic range. The proposed method outperforms state-of-the-art methods in terms of RMSE, SSIM, SRE, and SAM index, as given in Table~\ref{quan_tbl}.  Though there is minor degradation in terms of PSNR, we observe slight improvement in SRE which is a more reliable metric in satellite image processing~\cite{lanaras2018super}. Further, the percentage improvement of S2A over ALERT is as high as 9.4\% in RMSE and 8.1\% in SAM. The improvement over baseline is considerable given the fact that there exists similar gain between second and third best approach in the dataset under study. 

\begin{table}[t]
\begin{center}
\resizebox{0.5\textwidth}{!}{%
\begin{tabular}{|l|l|l|l|l|l|}
\hline
Method   & RMSE                         & SSIM(\%)                     & SRE(dB)                      & PSNR(dB)                     & SAM(deg)                    \\
\hline
\hline
AeroGAN~\cite{rangnekar2017aerial}   & 21.62                        & 86.03                        & 44.62                        & 36.50                        & 12.15                       \\
DSen2~\cite{lanaras2018super}     & 14.14                        & 93.85                        & 50.04                        & 41.94                        & 7.88                        \\
DeepSWIR~\cite{rout2019deepswir}  & {\color[HTML]{3166FF} 13.75} & {\color[HTML]{3166FF} 94.02} & {\color[HTML]{3166FF} 50.35} & {\color[HTML]{3166FF} 42.27} & {\color[HTML]{3166FF} 7.66} \\
ALERT~\cite{rout2020alert}     & {\color[HTML]{32CB00} 12.97} & {\color[HTML]{32CB00} 94.54} & {\color[HTML]{32CB00} 50.81} & {\color[HTML]{CB0000} 42.80} & {\color[HTML]{32CB00} 7.48} \\
S2A (ours) & {\color[HTML]{CB0000} 11.74} & {\color[HTML]{CB0000} 95.08} & {\color[HTML]{CB0000} 50.83} & {\color[HTML]{32CB00} 42.76} & {\color[HTML]{CB0000} 6.87} \\
\hline
\end{tabular}
} 
\end{center}
\caption{Quantitative analysis on LISS-3. First, second and third methods are highlighted as red, green and blue, respectively.} 
\label{quan_tbl}
\end{table}

\begin{figure}[t]
	\centering
	\includegraphics[width=0.75\columnwidth]{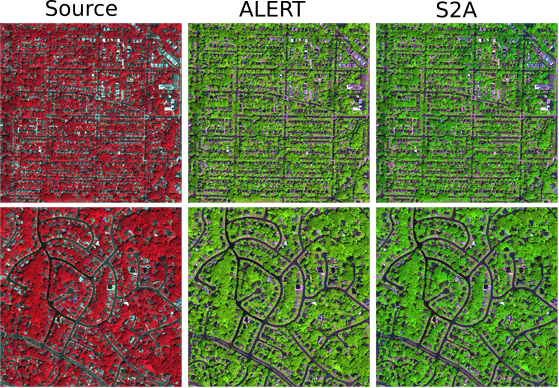}
	\caption{Analysis of multi-sensor (12x) WV-2 SWIR synthesis (Target domain: synthesized SWIR (R), NIR (G) and R (B)).}
	\label{wv2_alert_s2a}
\end{figure}

\subsection{Qualitative Analysis}
\subsubsection{Multi-Temporal}
Though the training data is acquired on Apr 07, 2018 and Jan 19, 2018, it is interesting to see how well the trained model generalizes to a scene acquired at a different time, Nov 17, 2017. Particularly intriguing is the phenomena of inverting features pre- and post-monsoon. As a consequence most of the static methods that do not use concurrent coarse resolution band~\cite{rangnekar2017aerial,rout2019deepswir,rout2020alert} fail to capture these dynamic relationships. Moreover, the proposed S2A generalizes reasonably well to multi-temporal data due to the use of attention map, as shown in Figure~\ref{mul_temp}. In this scenario, S2A captures certain radiometric information through spatial attention map from discriminator which has access to the concurrent coarse resolution band. 

\subsubsection{Multi-Resolution}
Though DSen2~\cite{lanaras2018super} also uses concurrent coarse resolution SWIR, it is not efficient in \textit{extrapolating} across multiple resolutions. To put more succinctly, we super-resolve LISS-3 SWIR to LISS-4 resolution (4x) using proposed S2A and DSen2~\cite{lanaras2018super}. While DSen2 relies heavily on the actual LISS-3 SWIR band, S2A uses it just to get a glimpse of relevant parts of a scene. Over dependency of DSen2 on \mbox{LISS-3} SWIR results in blurry edges as can be inferred from Figure~\ref{mul_res}(a). One can find shades of different colors, e.g., pinkish at these blurry edges even though all bands are registered properly. By looking at other edges it can be inferred that the shades are not due to mis-registration, but mainly due to inherent blur in band synthesis using DSen2~\cite{lanaras2018super}. However, S2A synthesizes sharper SWIR due to careful combination of existing high resolution bands of LISS-4 and spatial attention map from coarse resolution SWIR of LISS-3.

\begin{figure}[t]
	\centering
	\includegraphics[width=0.75\columnwidth]{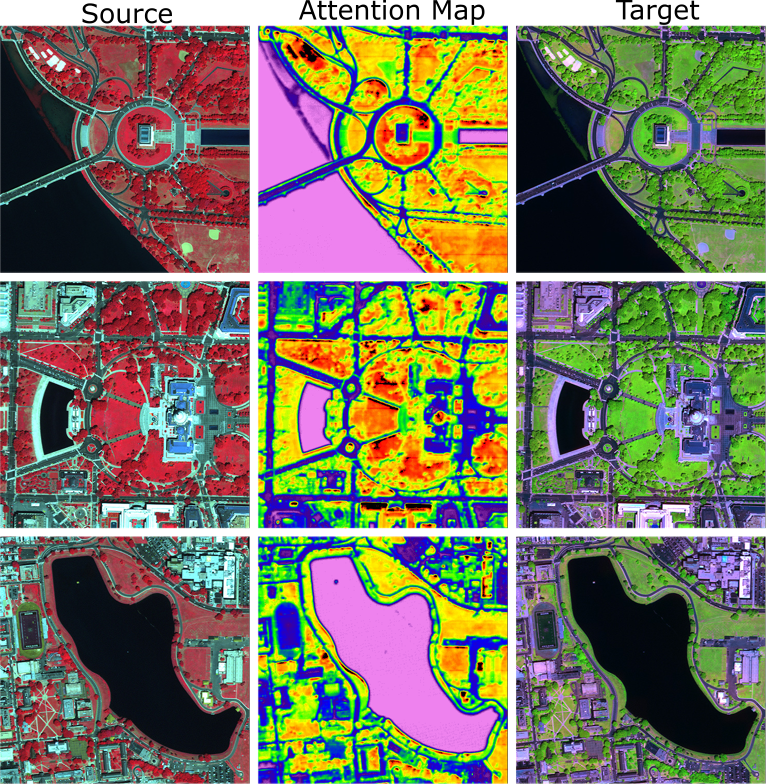}
	\caption{Spatial attention maps in multi-sensor band synthesis.}
	\label{sp_attn_iao}
\end{figure}

\subsubsection{Multi-Sensor}
To this end, we analyzed the performance on LISS-3 and LISS-4 imagery onboard R2A. However, to understand how well the proposed model generalizes to very high resolution images over a completely different physical landscape, we intend to synthesize SWIR at the resolution of WV-2, i.e, 1.84m GSD. Even in the absence of coarse-resolution SWIR in WV-2, S2A can synthesize it at the resolution of existing multi-spectral bands. While the availability of coarse resolution band benefits super-resolution (ref. Figure~\ref{mul_temp}), it certainly is not a limiting factor in the context of band synthesis. This is illustrated in Figure~\ref{wv2_alert_s2a} where high resolution SWIR synthesized by both S2A and ALERT are comparable qualitatively. Note that ALERT does not require coarse SWIR as it is a purely band-to-band translation approach. If the proposed model were over dependent on coarse SWIR, it would have produced a band similar to NIR which was used to extract spatial attention map in the absence of low resolution SWIR. But clearly S2A synthesized a band which is more like SWIR and less like NIR as evident from Figure~\ref{wv2_alert_s2a}. The mere fact that the false color composite containing synthesized SWIR is similar to ALERT supports this hypothesis. In Figure~\ref{sp_attn_iao}, we show that the model attends to relevant attributes during synthesis of WV-2 SWIR at 1.84m GSD.

\begin{figure}[t]
	\centering
	\includegraphics[width=0.75\columnwidth]{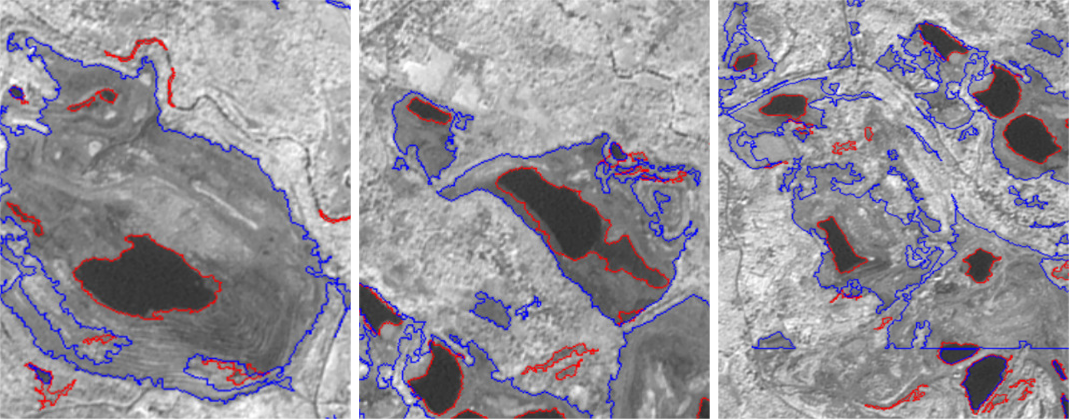}
	\caption{Wetland boundary detection using synthesized high resolution SWIR (Red) and existing high resolution NIR (Blue).}
	\label{wetland}
\end{figure}

\subsection{Application of Synthesized Band}
\textbf{Wetland Delineation:} A potential application of synthesized high resolution SWIR band is found in wetland delineation. Due to very low reflectance in the SWIR part of the spectrum, it can be beneficial in segregating wetland resources. Since wetland delineation is not the primary focus of this study, we use a naive thresholding approach to map wetlands. As shown in Figure~\ref{wetland}, synthesized SWIR band successfully delineates majority of wetlands including difficult to identify narrow wetlands. On the contrary, the use of NIR at the resolution of SWIR finds many false positives. It is to be noted that absorption due to water molecules is an inherent property of SWIR band. However, the absence of highly spatially resolved SWIR is a major issue in such applications. In this study, we merely provided a mechanism to super-resolve multi-spectral band that is shown to preserve required spectral characteristics. 

\textbf{Modified Normalized Difference Water Index (MNDWI): } In order to assess the scientific fidelity of super-resolved band, we analyze the MNDWI~\cite{han2005study} metric of various methods using equation~(\ref{ndwi_eq}). To quantify the results, we rely on Intersection over Union (IoU) measure commonly used in semantic segmentation~\cite{long2015fully}. 
\begin{equation}
    MNDWI = \frac{G-SWIR}{G+SWIR}
    \label{ndwi_eq}
\end{equation}

\begin{figure}[t]
	\centering
	\includegraphics[width=0.75\columnwidth]{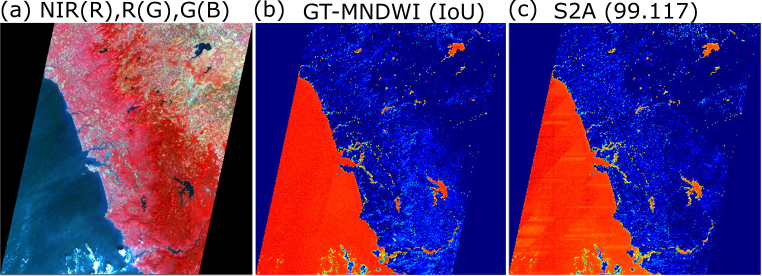}
	\caption{Qualitative comparison of MNDWI.}
	\label{ndwi}
\end{figure}

\begin{table}[t]
\begin{center}
\resizebox{0.5\textwidth}{!}{%
\begin{tabular}{|l|l|l|l|l|l|}
\hline
Method & AeroGAN~\cite{rangnekar2017aerial} & DSen2~\cite{lanaras2018super}  & DeepSWIR~\cite{rout2019deepswir} & ALERT~\cite{rout2020alert}  & S2A (ours) \\
\hline
\hline
IoU    & 97.181  & {\color[HTML]{3166FF} 98.891}  & 98.853   & {\color[HTML]{32CB00} 99.066} & {\color[HTML]{FE0000} 99.117} \\
\hline
\end{tabular}
} 
\end{center}
\caption{Quantitative comparison of MNDWI. S2A performs favorably against state-of-the-art methods.} 
\label{ndwi_tbl}
\end{table}
After quantitatively analyzing the IoU measure in Table~\ref{ndwi_tbl}, we compare MNDWI maps of Ground Truth (GT) and S2A qualitatively in Figure~\ref{ndwi}. Among the compared methods, S2A achieves state-of-the-art result.

\begin{figure}[t]
	\centering
	\includegraphics[width=0.75\columnwidth]{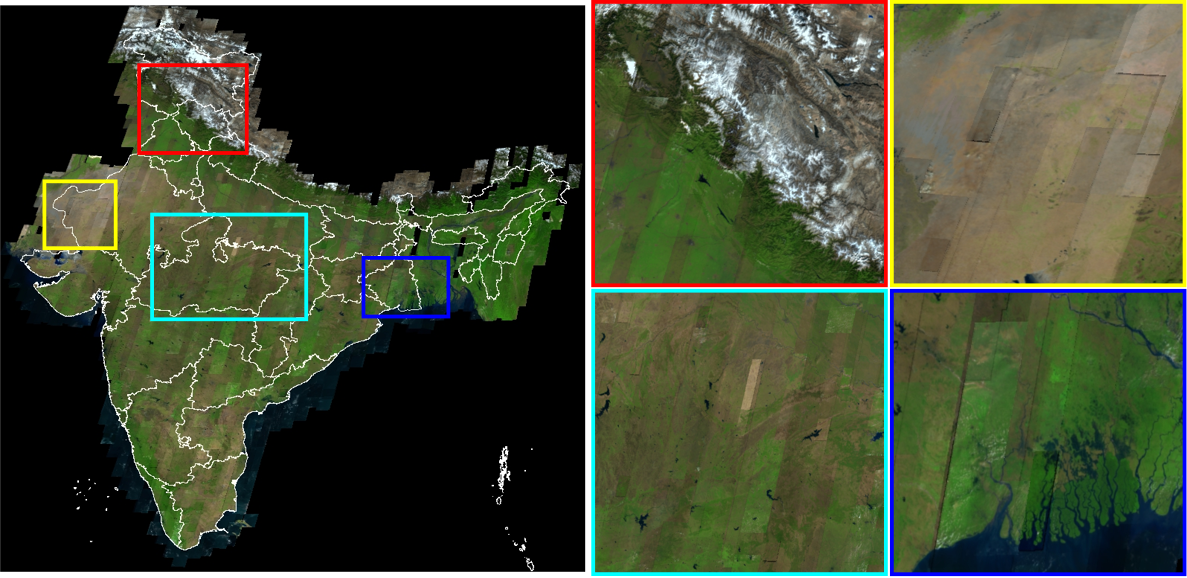}
	\caption{Large area mosaic of synthesized SWIR (R), NIR (G), and R (B) without relative radiometric normalization.  Highlighted portions include hilly, desert, main land, and coastal region over Indian land terrains (stretched for visualization).}
	\label{l4_mosaic}
\end{figure}

\textbf{Value Added Product:} Using the proposed methodology we synthesize SWIR at 5m GSD for approximately 4000 LISS-4 scenes covering various target attributes. Figure~\ref{l4_mosaic} illustrates the false color composite over Indian land terrains. It is to be noted that the merged products include multi-temporal scenes that may appear as radiometric imbalance in the absence of normalization. To preserve scientific fidelity, we show various land terrains without normalization in Figure~\ref{l4_mosaic}.

\section{Conclusion}
\label{conc}
In this study, we primarily focused on spatial and spectral attention with respect to multi-spectral band synthesis. We formulated the ill-posed super-resolution problem as conditional band synthesis, and explored various state-of-the-art methods in this regard. In the process, we developed a WGAN based adversarial framework encapsulating spatial and Laplacian spectral channel attention. Further, we introduced two new cost functions, namely spatial attention loss and domain adaptation loss to improve scientific fidelity in the context of multi-spectral band synthesis. In addition, we critically analyzed the qualitative and quantitative results among the compared methods. Our experiments on three different satellite imagery: LISS-3, LISS-4, and WorldView-2 over isolated physical landscape demonstrated generalization ability of the proposed method. Finally, we discussed several real world applications of the synthesized band that include wetland delineation, water masking, and additional value product generation. Based on our experiments, we report favorable results in multi-spectral high resolution band synthesis. Though we showcased the efficacy of spatial attention and domain adaptation loss in remote sensing, we believe these methods can also be employed in non-remote sensing imagery. We defer discussion on these questions to future work. 

{\small
\bibliographystyle{ieee_fullname}
\bibliography{egbib}

\begin{thebibliography}{10}\itemsep=-1pt

\bibitem{anwar2019densely}
Saeed Anwar and Nick Barnes.
\newblock Densely residual laplacian super-resolution.
\newblock {\em arXiv preprint arXiv:1906.12021}, 2019.

\bibitem{arbelaez2010contour}
Pablo Arbelaez, Michael Maire, Charless Fowlkes, and Jitendra Malik.
\newblock Contour detection and hierarchical image segmentation.
\newblock {\em IEEE transactions on pattern analysis and machine intelligence},
  33(5):898--916, 2010.

\bibitem{arjovsky2017wasserstein}
Martin Arjovsky, Soumith Chintala, and L{\'e}on Bottou.
\newblock Wasserstein gan.
\newblock {\em arXiv preprint arXiv:1701.07875}, 2017.

\bibitem{bahdanau2014neural}
Dzmitry Bahdanau, Kyunghyun Cho, and Yoshua Bengio.
\newblock Neural machine translation by jointly learning to align and
  translate.
\newblock {\em arXiv preprint arXiv:1409.0473}, 2014.

\bibitem{bastidas2019channel}
Alexei~A Bastidas and Hanlin Tang.
\newblock Channel attention networks.
\newblock In {\em Proceedings of the IEEE Conference on Computer Vision and
  Pattern Recognition Workshops}, pages 0--0, 2019.

\bibitem{beaulieu2018deep}
Mario Beaulieu, Samuel Foucher, Dan Haberman, and Colin Stewart.
\newblock Deep image-to-image transfer applied to resolution enhancement of
  sentinel-2 images.
\newblock In {\em IGARSS 2018-2018 IEEE International Geoscience and Remote
  Sensing Symposium}, pages 2611--2614. IEEE, 2018.

\bibitem{bevilacqua2012low}
Marco Bevilacqua, Aline Roumy, Christine Guillemot, and Marie~Line
  Alberi-Morel.
\newblock Low-complexity single-image super-resolution based on nonnegative
  neighbor embedding.
\newblock 2012.

\bibitem{chen2016attention}
Liang-Chieh Chen, Yi Yang, Jiang Wang, Wei Xu, and Alan~L Yuille.
\newblock Attention to scale: Scale-aware semantic image segmentation.
\newblock In {\em Proceedings of the IEEE conference on computer vision and
  pattern recognition}, pages 3640--3649, 2016.

\bibitem{dong2015image}
Chao Dong, Chen~Change Loy, Kaiming He, and Xiaoou Tang.
\newblock Image super-resolution using deep convolutional networks.
\newblock {\em IEEE transactions on pattern analysis and machine intelligence},
  38(2):295--307, 2015.

\bibitem{emami2019spa}
Hajar Emami, Majid~Moradi Aliabadi, Ming Dong, and Ratna~Babu Chinnam.
\newblock Spa-gan: Spatial attention gan for image-to-image translation.
\newblock {\em arXiv preprint arXiv:1908.06616}, 2019.

\bibitem{goodfellow2014generative}
Ian Goodfellow, Jean Pouget-Abadie, Mehdi Mirza, Bing Xu, David Warde-Farley,
  Sherjil Ozair, Aaron Courville, and Yoshua Bengio.
\newblock Generative adversarial nets.
\newblock In {\em Advances in neural information processing systems}, pages
  2672--2680, 2014.

\bibitem{gulrajani2017improved}
Ishaan Gulrajani, Faruk Ahmed, Martin Arjovsky, Vincent Dumoulin, and Aaron~C
  Courville.
\newblock Improved training of wasserstein gans.
\newblock In {\em Advances in neural information processing systems}, pages
  5767--5777, 2017.

\bibitem{han2005study}
XU Han-Qiu.
\newblock A study on information extraction of water body with the modified
  normalized difference water index (mndwi).
\newblock {\em Journal of remote sensing}, 5:589--595, 2005.

\bibitem{haut2018new}
Juan~Mario Haut, Ruben Fernandez-Beltran, Mercedes~E Paoletti, Javier Plaza,
  Antonio Plaza, and Filiberto Pla.
\newblock A new deep generative network for unsupervised remote sensing
  single-image super-resolution.
\newblock {\em IEEE Transactions on Geoscience and Remote Sensing},
  56(11):6792--6810, 2018.

\bibitem{huang2015single}
Jia-Bin Huang, Abhishek Singh, and Narendra Ahuja.
\newblock Single image super-resolution from transformed self-exemplars.
\newblock In {\em Proceedings of the IEEE conference on computer vision and
  pattern recognition}, pages 5197--5206, 2015.

\bibitem{huang2017single}
Ningbo Huang, Yong Yang, Junjie Liu, Xinchao Gu, and Hua Cai.
\newblock Single-image super-resolution for remote sensing data using deep
  residual-learning neural network.
\newblock In {\em International Conference on Neural Information Processing},
  pages 622--630. Springer, 2017.

\bibitem{isola2017image}
Phillip Isola, Jun-Yan Zhu, Tinghui Zhou, and Alexei~A Efros.
\newblock Image-to-image translation with conditional adversarial networks.
\newblock In {\em Proceedings of the IEEE conference on computer vision and
  pattern recognition}, pages 1125--1134, 2017.

\bibitem{jolicoeur2018relativistic}
Alexia Jolicoeur-Martineau.
\newblock The relativistic discriminator: a key element missing from standard
  gan.
\newblock {\em arXiv preprint arXiv:1807.00734}, 2018.

\bibitem{Kim_2016_CVPR}
Jiwon Kim, Jung Kwon~Lee, and Kyoung Mu~Lee.
\newblock Accurate image super-resolution using very deep convolutional
  networks.
\newblock In {\em The IEEE Conference on Computer Vision and Pattern
  Recognition (CVPR)}, June 2016.

\bibitem{Lai_2017_CVPR}
Wei-Sheng Lai, Jia-Bin Huang, Narendra Ahuja, and Ming-Hsuan Yang.
\newblock Deep laplacian pyramid networks for fast and accurate
  super-resolution.
\newblock In {\em The IEEE Conference on Computer Vision and Pattern
  Recognition (CVPR)}, July 2017.

\bibitem{lanaras2018super}
Charis Lanaras, Jos{\'e} Bioucas-Dias, Silvano Galliani, Emmanuel Baltsavias,
  and Konrad Schindler.
\newblock Super-resolution of sentinel-2 images: Learning a globally applicable
  deep neural network.
\newblock {\em ISPRS Journal of Photogrammetry and Remote Sensing},
  146:305--319, 2018.

\bibitem{Ledig_2017_CVPR}
Christian Ledig, Lucas Theis, Ferenc Huszar, Jose Caballero, Andrew Cunningham,
  Alejandro Acosta, Andrew Aitken, Alykhan Tejani, Johannes Totz, Zehan Wang,
  and Wenzhe Shi.
\newblock Photo-realistic single image super-resolution using a generative
  adversarial network.
\newblock In {\em The IEEE Conference on Computer Vision and Pattern
  Recognition (CVPR)}, July 2017.

\bibitem{lei2017super}
Sen Lei, Zhenwei Shi, and Zhengxia Zou.
\newblock Super-resolution for remote sensing images via local--global combined
  network.
\newblock {\em IEEE Geoscience and Remote Sensing Letters}, 14(8):1243--1247,
  2017.

\bibitem{long2015fully}
Jonathan Long, Evan Shelhamer, and Trevor Darrell.
\newblock Fully convolutional networks for semantic segmentation.
\newblock In {\em Proceedings of the IEEE conference on computer vision and
  pattern recognition}, pages 3431--3440, 2015.

\bibitem{luo2017video}
Yimin Luo, Liguo Zhou, Shu Wang, and Zhongyuan Wang.
\newblock Video satellite imagery super resolution via convolutional neural
  networks.
\newblock {\em IEEE Geoscience and Remote Sensing Letters}, 14(12):2398--2402,
  2017.

\bibitem{mirza2014conditional}
Mehdi Mirza and Simon Osindero.
\newblock Conditional generative adversarial nets.
\newblock {\em arXiv preprint arXiv:1411.1784}, 2014.

\bibitem{mnih2014recurrent}
Volodymyr Mnih, Nicolas Heess, Alex Graves, et~al.
\newblock Recurrent models of visual attention.
\newblock In {\em Advances in neural information processing systems}, pages
  2204--2212, 2014.

\bibitem{mo2018instagan}
Sangwoo Mo, Minsu Cho, and Jinwoo Shin.
\newblock Instagan: Instance-aware image-to-image translation.
\newblock {\em arXiv preprint arXiv:1812.10889}, 2018.

\bibitem{park2018srfeat}
Seong-Jin Park, Hyeongseok Son, Sunghyun Cho, Ki-Sang Hong, and Seungyong Lee.
\newblock Srfeat: Single image super-resolution with feature discrimination.
\newblock In {\em Proceedings of the European Conference on Computer Vision
  (ECCV)}, pages 439--455, 2018.

\bibitem{radford2015unsupervised}
Alec Radford, Luke Metz, and Soumith Chintala.
\newblock Unsupervised representation learning with deep convolutional
  generative adversarial networks.
\newblock {\em arXiv preprint arXiv:1511.06434}, 2015.

\bibitem{rangnekar2017aerial}
Aneesh Rangnekar, Nilay Mokashi, Emmett Ientilucci, Christopher Kanan, and
  Matthew Hoffman.
\newblock Aerial spectral super-resolution using conditional adversarial
  networks.
\newblock {\em arXiv preprint arXiv:1712.08690}, 2017.

\bibitem{rout2020alert}
Litu Rout.
\newblock Alert: Adversarial learning with expert regularization using tikhonov
  operator for missing band reconstruction.
\newblock {\em IEEE Transactions on Geoscience and Remote Sensing}, 2020.

\bibitem{rout2019deepswir}
Litu Rout, Yatharath Bhateja, Ankur Garg, Indranil Mishra, S~Manthira Moorthi,
  and Debjyoti Dhar.
\newblock Deepswir: A deep learning based approach for the synthesis of
  short-wave infrared band using multi-sensor concurrent datasets.
\newblock {\em arXiv preprint arXiv:1905.02749}, 2019.

\bibitem{sajjadi2017enhancenet}
Mehdi~SM Sajjadi, Bernhard Scholkopf, and Michael Hirsch.
\newblock Enhancenet: Single image super-resolution through automated texture
  synthesis.
\newblock In {\em Proceedings of the IEEE International Conference on Computer
  Vision}, pages 4491--4500, 2017.

\bibitem{Shi_2016_CVPR}
Wenzhe Shi, Jose Caballero, Ferenc Huszar, Johannes Totz, Andrew~P. Aitken, Rob
  Bishop, Daniel Rueckert, and Zehan Wang.
\newblock Real-time single image and video super-resolution using an efficient
  sub-pixel convolutional neural network.
\newblock In {\em The IEEE Conference on Computer Vision and Pattern
  Recognition (CVPR)}, June 2016.

\bibitem{vaswani2017attention}
Ashish Vaswani, Noam Shazeer, Niki Parmar, Jakob Uszkoreit, Llion Jones,
  Aidan~N Gomez, {\L}ukasz Kaiser, and Illia Polosukhin.
\newblock Attention is all you need.
\newblock In {\em Advances in neural information processing systems}, pages
  5998--6008, 2017.

\bibitem{wang2017residual}
Fei Wang, Mengqing Jiang, Chen Qian, Shuo Yang, Cheng Li, Honggang Zhang,
  Xiaogang Wang, and Xiaoou Tang.
\newblock Residual attention network for image classification.
\newblock In {\em Proceedings of the IEEE Conference on Computer Vision and
  Pattern Recognition}, pages 3156--3164, 2017.

\bibitem{wang2018esrgan}
Xintao Wang, Ke Yu, Shixiang Wu, Jinjin Gu, Yihao Liu, Chao Dong, Yu Qiao, and
  Chen Change~Loy.
\newblock Esrgan: Enhanced super-resolution generative adversarial networks.
\newblock In {\em Proceedings of the European Conference on Computer Vision
  (ECCV)}, pages 0--0, 2018.

\bibitem{xu2015show}
Kelvin Xu, Jimmy Ba, Ryan Kiros, Kyunghyun Cho, Aaron Courville, Ruslan
  Salakhudinov, Rich Zemel, and Yoshua Bengio.
\newblock Show, attend and tell: Neural image caption generation with visual
  attention.
\newblock In {\em International conference on machine learning}, pages
  2048--2057, 2015.

\bibitem{yang2010image}
Jianchao Yang, John Wright, Thomas~S Huang, and Yi Ma.
\newblock Image super-resolution via sparse representation.
\newblock {\em IEEE transactions on image processing}, 19(11):2861--2873, 2010.

\bibitem{yuhas1992discrimination}
Roberta~H Yuhas, Alexander~FH Goetz, and Joe~W Boardman.
\newblock Discrimination among semi-arid landscape endmembers using the
  spectral angle mapper (sam) algorithm.
\newblock 1992.

\bibitem{zagoruyko2016paying}
Sergey Zagoruyko and Nikos Komodakis.
\newblock Paying more attention to attention: Improving the performance of
  convolutional neural networks via attention transfer.
\newblock {\em arXiv preprint arXiv:1612.03928}, 2016.

\bibitem{zhang2018self}
Han Zhang, Ian Goodfellow, Dimitris Metaxas, and Augustus Odena.
\newblock Self-attention generative adversarial networks.
\newblock {\em arXiv preprint arXiv:1805.08318}, 2018.

\bibitem{Zhang_2018_CVPR}
Yulun Zhang, Yapeng Tian, Yu Kong, Bineng Zhong, and Yun Fu.
\newblock Residual dense network for image super-resolution.
\newblock In {\em The IEEE Conference on Computer Vision and Pattern
  Recognition (CVPR)}, June 2018.

\bibitem{zhu2017unpaired}
Jun-Yan Zhu, Taesung Park, Phillip Isola, and Alexei~A Efros.
\newblock Unpaired image-to-image translation using cycle-consistent
  adversarial networks.
\newblock In {\em Proceedings of the IEEE international conference on computer
  vision}, pages 2223--2232, 2017.

\end{thebibliography}
}

\end{document}